\documentclass[twocolumn,aps,showpacs,prl]{revtex4}

\usepackage{amsmath}
\usepackage{amssymb}
\usepackage{psfrag}
\usepackage[dvips]{graphicx}

\begin{document}
\title{Better Late than Never: Information Retrieval from Black Holes}
\author{Samuel L.\ Braunstein}
\author{Stefano Pirandola}
\affiliation{Department of Computer Science, University of York,
York YO10 5GH, United Kingdom}
\author{Karol \.{Z}yczkowski}
\affiliation{Institute of Physics, Jagiellonian University,
30-059 Krakow, Poland}
\affiliation{Center for Theoretical Physics,
Polish Academy of Science, 02-668 Warszawa, Poland}
\date{Received 17 April 2012}

\begin{abstract}
We show that, in order to preserve the equivalence principle until late
times in unitarily evaporating black holes, the thermodynamic entropy of
a black hole must be primarily entropy of entanglement across the event
horizon. For such black holes, we show that the information entering
a black hole becomes encoded in correlations within a tripartite
quantum state, the quantum analogue of a one-time pad, and is only
decoded into the outgoing radiation very late in the evaporation.
This behavior generically describes the unitary evaporation of highly
entangled black holes and requires no specially designed evolution.
Our work suggests the existence of a matter-field sum rule for any
fundamental theory.
\end{abstract}


\maketitle

{\it Black hole evaporation as tunneling}.---%
Although pair creation provides the conventional heuristic picture of
the microscopic process by which a black hole evaporates
\cite{Hawking75}, it has come under increasing suspicion due to intrinsic
difficulties. In particular, pair creation necessarily requires the
dimensionality of the interior Hilbert space of a black hole to
be increasing while simultaneously its physical size is
decreasing \cite{Nikolic,B10}.

By contrast, quantum tunneling, which operates by moving quantum
subsystems across the classically forbidden barrier of the event
horizon, naturally avoids this difficulty \cite{B10}. Furthermore,
quantum tunneling invites an elegant Hilbert space description of the
evaporation process across event horizons \cite{B10}: We start
with the standard decomposition of a black hole Hilbert space into a
tensor product between the interior (int) and exterior (ext) by
${\cal H}_{\text{int}}\otimes{\cal H}_{\text{ext}}$ \cite{Hawking76}
and note that an event horizon's tensor product structure in no
way implies that its spatial location cannot be fuzzy \cite{B10}.

Tunneling now operates \cite{B10} by selecting some subsystem from
the black hole interior and moving it to the exterior
${\cal H}_{\text{int}}\rightarrow{\cal H}_B\otimes {\cal H}_R$ by
\begin{equation}
|i\rangle_{\text{int}}\rightarrow (U|i\rangle)_{BR}, \label{Umodel}
\end{equation}
where $U$ denotes the unitary process that might be thought of as
``selecting'' the subsystem to eject, $|i\rangle$ is the initial
state of the black hole interior, $B$ denotes the {\it reduced size\/}
subsystem corresponding to the remaining interior after evaporation,
and $R$ denotes the subsystem that escapes as
radiation \cite{B10,Page93,Hayden07}.

Equation~(\ref{Umodel}) has been used before to study black hole
evaporation \cite{B10,Page93,Hayden07}; however, with the exception
of Ref.~[\onlinecite{B10}], it has not been used as a process
associated with any
underlying physical mechanism. Indeed, Ref.~[\onlinecite{B10}] showed that
the symmetries implicit in this equation, in conjunction with global 
conservation laws for the no-hair quantities (energy, charge, and 
angular momentum), suffice to completely determine black hole tunneling 
probabilities for any black hole and particle type, reproducing and 
even extending the predictions of field theory on curved spacetime. 
This work therefore strongly supports Eq.~(\ref{Umodel}) as a pertinent
microscopic formulation of unitary black hole evaporation. Its
implications for the retrieval of information about in-fallen
matter will be further studied here.

{\it Dynamical evaporation with entanglement}.---%
It is now well accepted that entanglement across boundaries is
generic \cite{Eisert09}. Therefore, our key point of departure from
previous work \cite{B10,Page93,Hayden07,Smolin,me} will be to allow for
entanglement across the event horizon. Incorporated into the evaporative
dynamics of Eq.~(\ref{Umodel}), but making no assumption of how much or
how little transevent horizon entanglement there may be, this
entanglement gives
\begin{equation}
\sum_{i=1}^N\sqrt{p_i}\, |i\rangle_{\text{int}}\otimes|i\rangle_{\text{ext}}
\rightarrow \sum_{i=1}^N \sqrt{p_i}\, (U|i\rangle)_{{BR}}\otimes
|i\rangle_{\text{ext}}.
\label{entUmodel}
\end{equation}

Now, the nature of the black hole as a compact object of a given mass 
constrains any interior evolution to only access an effectively {\it 
finite\/} dimensional Hilbert space \cite{B10}. Quantities defined within
(the support of) this finite Hilbert space will similarly be finite,
including, for example, any von Neumann entropies, measures of
entanglement, etc. Indeed, it has been argued \cite{Susskind93}
that the dimensionality for the initial black hole Hilbert space should
be well approximated by the {\it thermodynamic\/} entropy
$S_{\text{BH}}={\cal A}/(4\ln 2)$ for a black hole of area ${\cal A}$,
giving a dimensionality
$N\equiv\text{dim}({\cal H}_\text{int})= BR= 2^{S_{\text{BH}}}$,
where we reuse subsystem labels for Hilbert space dimensionalities
and for later convenience we evaluate entropies using base-two logarithms.
We might say that the black hole interior comprises
$\log_2 N=S_{\text{BH}}$ qubits. (Throughout, the term ``qubits'' is used
merely as a unit of information content and does not literally imply 
a set of two-level systems.) That the number of qubits
initially within the black hole is well approximated by $S_{\text{BH}}$
is supported by the holographic principle \cite{tHooft93} and 
independently by the amount of Hawking radiation that would be
generated, consistent with energy conservation. 

Naively, to make quantitative predictions based on this description,
we would need to know the detailed dynamics $U$ within the black hole.
In fact, the behavior of information flow in a high-dimensional system 
under a specific unitary will be in excellent agreement with the Haar 
average over all unitaries acting on dimension $N$. This follows from
Levy's lemma \cite{Levy}, which states that the logarithm of the
probability of any such difference $\epsilon$ scales as $-N \epsilon^2$. 
For a stellar mass black hole, such dimensionalities $N$ must be at least 
${10^{10}}^{77}$, so even the smallest deviations from the average behavior 
should occur with vanishingly small probability. Numerical simulations
in even very low dimensions show this to be well supported, and similar
results are well known beyond black hole physics \cite{Mehta}. Thus, here
we replace the behavior of the specific unitary in Eq.~(\ref{entUmodel}) 
by the Haar average.

{\it Vanishing of transevent horizon entanglement}.---%
Moving to the average behavior allows one to
rigorously interpret the evaporative dynamics of a black hole in terms
of the properties of random quantum error correcting codes \cite{Hayden07}.
In this interpretation, one-half of an entangled state is encoded
into a larger Hilbert space via a random unitary encoding. Decoupling
theorems \cite{Abey06} tell us how much (how many qubits) of the encoded
state one must have access to, in principle, in order to reconstruct the
original unencoded state, including its entanglement. We derive a
generalized decoupling theorem and use it to address a broader set of
questions. (See the Supplemental Material in Ref.~[\onlinecite{SM}] for
proofs and a discussion of both quantum and classical decoupling
theorems.)

For example, for an entangled black hole evolving via
Eq.~(\ref{entUmodel}), this generalized decoupling theorem shows that,
for any positive number $c$, once
\begin{equation}
\log_2 R=\frac{1}{2}S_{\text{BH}}+\frac{1}{2}H^{(1/2)}(\rho_{\text{ext}})+c
\label{entVanishing}
\end{equation}
qubits have radiated away, the transevent horizon entanglement
will have vanished, appearing instead, with virtually unit fidelity
(at least $1-2^{-c}$), as entanglement between the external neighborhood
and radiation. Here, the entropy of entanglement is quantified by a
R\'enyi entropy
$H^{(q)}(\rho) \equiv \log_2({\text{tr}}\, \rho^q)/(1-q)$ with $q$
of order unity, for the reduced density matrix of the (ext) state
$\rho_{\text{ext}}=\sum_{i=1}^N p_i |i\rangle_{\text{ext}}\,
{}_{\text{ext}} \langle i|$ neighboring the event horizon.

{\it Entanglement and the equivalence principle}.---%
We will now explicitly link the presence of transevent horizon 
entanglement with the equivalence principle. Specifically, the 
equivalence principle is expected to be preserved for black holes 
larger than the Planck scale. We will argue below that the presence
of this entanglement must be similarly preserved until such scales.
We then use the tunneling dynamics to calculate the initial amount
of transevent horizon entanglement.
  
We start by recalling the equivalence principle, 
which tells us that a freely falling observer sees no {\it local\/} 
effects due to gravity. Applied to black holes, it has been
argued \cite{Susskind93} that the equivalence principle implies that
an observer freely falling past the event horizon would see no Hawking
radiation, only a zero temperature vacuum state---just as an
unaccelerated observer in flat spacetime. Now, the well-understood quantum
physics of condensed matter systems tells us that entanglement across
boundaries is generic in or near the ground state \cite{Eisert09}.
Furthermore, in axiomatic quantum field theory, entanglement across
boundaries for fields in their vacuum state is implicit in the
Reeh-Schlieder theorem \cite{Schlieder}. 
In the Supplemental Material \cite{SM}, we derive a lower bound for the
energy of a free scalar field when the quantum state is restricted
to have no entanglement across an arbitrary hypothetical boundary.
This disentanglement energy diverges as a power of the UV regulator
\cite{SM}, and hence is far above the vacuum state. Applied to black
holes, this means that the loss of entanglement across the event horizon
%
would force the quantum fields across it to be arbitrarily far from the
vacuum state---an {\it energetic curtain\/} would have descended around
the black hole \cite{SLB2009}---signaling
%
a manifest failure of the equivalence principle.

Next, we use the epoch for the loss of transevent horizon entanglement,
given by Eq.~(\ref{entVanishing}), to quantify how much transevent
horizon entanglement was in the initial black hole. Here, we rely on the
observation that a black hole's size may be directly quantified by its 
area or, equivalently, its entropy. 
For black holes in the latter stages of evaporation via
Eq.~(\ref{entUmodel}), their entropy is well approximated by
$S_{\text{BH}} - \log_2 R$ \cite{SM}. Therefore, an evaporating
black hole can be said to approach the Planck scale (see
Ref.~[\onlinecite{SM}] for a detailed discussion) when, to high
precision, $\log_2 R\approx S_{\text{BH}}$. From Eq.~(\ref{entVanishing}),
preserving entanglement until such late times implies that
\begin{equation}
H^{(1/2)}(\rho_{\text{ext}})\approx S_{\text{BH}}.
\label{entropyEnt}
\end{equation}
In other words, preserving transevent horizon entanglement up until
an evaporating black hole approaches the Planck scale requires that
its initial entanglement entropy be almost exactly its initial thermodynamic
entropy \cite{fnA}. This result is insensitive to where we place the
entry point to the Planck scale \cite{SM}. Furthermore, this equality
does not change when the quantum state of the matter that originally
collapsed to form the black hole is taken into account \cite{SM}.
Finally, we note that, for the special case where the transevent
horizon entangled state in Eq.~(\ref{entUmodel}) reduces to uniform
entanglement (where all nonzero probabilities are equal), 
Eq.~(\ref{entVanishing}) may be replaced by established results 
\cite{Hayden07}, allowing a straightforward check of our analysis
(see Ref.~[\onlinecite{SM}]).

{\it Incorporating in-fallen matter}.---%
Naively, one might expect the entropy of ordinary matter 
$S_{\text{matter}}$ that collapses to form a black hole to be a large 
fraction of a black hole's thermodynamic entropy. However, this is not 
the case: 't Hooft \cite{tHooft93} has shown that 
$S_{\text{matter}}\lesssim S_{\text{BH}}^{3/4}$. Thus, for anything but 
Planck scale black holes, the entropic contribution from in-fallen matter 
is negligible, $S_{\text{matter}}\lll S_{\text{BH}}$. This then raises
the question of when and in what fashion the information about the
in-fallen matter can be retrieved. The remainder of this Letter addresses
this question. We proceed from our result, Eq.~(\ref{entropyEnt}),
that a black hole's thermodynamic entropy is almost entirely entropy
of transevent horizon entanglement. In so doing, we need not further
appeal to the equivalence principle or the specific state of quantum
fields across the event horizon. 

We tag the matter by entanglement with some distant reference (ref)
subsystem \cite{me,Hayden07} and use the decoupling theorem to
track its flow. It is conventional to assume that there is no ``bleaching''
mechanism \cite{bleach} that can strip away any of the
information about the in-fallen matter as it collapses to form a
black hole. In that case, the exterior Hilbert space can contain
no information about it. Now, the {\it no-hiding theorem} \cite{me}
gives a unique description for a quantum state where information is
not available within some specific subsystem. No-hiding implies
that the quantum state of a newly formed black hole interior
(int) and its surroundings must have the form
\begin{equation}
\frac{1}{\sqrt{K}}\sum_{i=1}^K |i\rangle_{\text{ref}}\otimes
\sum_{j}\sqrt{p_j}\,(|i\rangle\otimes |j\rangle\oplus 0)_{\text{int}}
\otimes|j\rangle_{\text{ext}}, \tag{5a}\label{entangledI}
\end{equation}
up to overall int-local and ext-local unitaries. Here, $\oplus\, 0$ means
we pad unused dimensions of the interior space by zero vectors \cite{me},
and $\log_2 K\equiv S_{\text{matter}}$ is the number of qubits
describing the quantum state of the matter collapsing to form the black
hole.  

Applying the dynamics of Eq.~(\ref{Umodel}) to our entangled black hole,
in the presence of in-fallen matter, gives
\begin{equation}
\rightarrow
\frac{1}{\sqrt{K}}\sum_{i=1}^K |i\rangle_{\text{ref}}\otimes
\!\sum_{j}\sqrt{p_j}\,[U(|i\rangle\otimes |j\rangle\oplus 0)]_{BR}
\otimes|j\rangle_{\text{ext}}.
\tag{5b}\label{entangledF}
\end{equation}

{\it Information retrieval from entangled black holes}.---%
\setcounter{equation}{5}%
We now apply our generalized decoupling theorem to the
evaporative dynamics of Eq.~(5). In order to state our
results, it will be convenient to roughly quantify the number of
unentangled (pure) qubits within the initial black hole state
in Eq.~(\ref{entangledI}); we define this ``excess'' as
\begin{equation}
\chi^{(q)}\equiv S_{\text{BH}} - S_{\text{matter}}
 -H^{(q)}(\rho_{\text{ext}})\ge 0.  \label{chiq}
\end{equation}
Note that Eq.~(\ref{entropyEnt}) implies
$\chi^{(1/2)} \lll S_{\text{BH}}$.

We now summarize the results about information encoding and retrieval. 
Since, in each application of the theorem, an independent dummy variable
appears [$c$ in Eq.~(\ref{entVanishing})] that is dwarfed by
other entropies, here we omit reference to them (the complete 
statements can be found in Ref.~[\onlinecite{SM}]).

Thermalization: Initially, one might suppose that any 
in-fallen matter would be well
within the interior of the black hole, far inside the event horizon,
and so would not be selected by $U$ to participate in tunneling across
this boundary. Only after the black hole had sufficiently ``scrambled''
the internal states (after what might be called the global thermalization
time \cite{Hayden07} for the black hole) would the subsystem encoding
the state of the in-fallen matter be accessible for selection and
ejection by tunneling \cite{fnB}. Note that estimates of scrambling
times vary. Some recent analyses suggest that black holes are fast 
scramblers \cite{Hayden07,Sekino08} (with the scrambling time being little 
more than the time for a single Hawking photon to evaporate), whereas 
other estimates are slow \cite{Giddings07}. 

Encoding: During the global thermalization time and for the next
$\frac{1}{2}\chi^{(1/2)}$ qubits radiated, all the information
about the state of the in-fallen matter is encoded with virtually
unit fidelity within the black hole interior. For the next
$S_{\text{matter}}+\frac{1}{2}(\chi^{(2)}-\chi^{(1/2)})$ qubits
radiated, this information becomes encoded into the tripartite
correlations of a {\it quantum one-time pad} \cite{me} among the black
hole interior, the external neighborhood, and the radiation. In other
words, it is the evaporation via tunneling (across the event horizon)
that encodes the information as tripartite entanglement. After encoding
and until the last $S_{\text{matter}}+\frac{1}{2} \chi^{(2)}$
qubits radiated the information remains within this quantum one-time pad;
it is inaccessible from any subsystem individually, but it is
accessible from any two of them. The quantum one-time pad is a random
quantum error correction code. The properties of such codes dictate the
size of subsystems one must have access to in order to be able to
reconstruct the original state of the in-fallen matter.

Decoding:
At this point in the evaporation process, entanglement within the 
black hole becomes so depleted that it can no longer contain the 
correlations of all the in-fallen matter. The final
$S_{\text{matter}}+\frac{1}{2} \chi^{(2)}$ qubits to be radiated marks
the start of information release into the radiation. From here until
the final $\frac{1}{2}\chi^{(1/2)}$ qubits radiated from the black hole, 
the full information about the in-fallen matter is decoded and becomes
available in the outgoing radiation for the first time. This decoding
takes the same amount of time as the encoding. Since typically
$\chi^{(2)}-\chi^{(1/2)} \lesssim O(1)$ and this 
quantity cannot be negative, the encoding or decoding occurs at roughly the 
radiation emission rate; recall, Hawking quanta typically carry around
one (qu)bit of thermal entropy. (See Ref.~[\onlinecite{SM}] for a
heuristic picture of the flow of information.) 

This completes our analysis of information retrieval for unitarily
evaporating highly entangled black holes.
As is evident from this
summary, decoding of the information about in-fallen matter is
very brief, occurring within the final and vanishingly small fraction
$S_{\text{matter}}/S_{\text{BH}}\lesssim O(S_{\text{BH}}^{-1/4})$
of a large black hole's lifetime (as measured in Hawking quanta
radiated). This is so late that its timing is unaffected by even
very long scrambling times \cite{Giddings07}. That said, within this
very brief epoch, decoding is also very slow, occurring at the
radiation emission rate; thus, information about the in-fallen
matter is decoded over the time scale required for
$S_{\text{matter}} \lesssim O(S_{\text{BH}}^{3/4})$ Hawking quanta
to evaporate. Because the number of qubits radiated during decoding
is so vast, essentially all the information has been retrieved long
before the black hole shrinks to the Planck scale.
Note that the time scales above follow from 't Hooft's
entropic bound \cite{tHooft93}; however, none of the mechanisms or
mathematical results in this Letter rely on this bound \cite{SM}.

{\it Discussion and wider implications}.---%
The application of information theoretic approaches to the physics of
black holes is relatively
new \cite{me,Nikolic,Page93,Hayden07,Smolin,B10,Susskind93}.
Here, we have shown that this approach offers a
description of black holes as highly entangled, with direct consequences
for the time course of information retrieval therefrom. 
This approach necessarily requires an explicit formulation of the 
microscopic evaporation process, which, here, we take to be quantum 
tunneling \cite{B10}. The analysis and the results are grounded
in black hole physics, and hence cannot be taken to apply to arbitrary
horizons, but the tunneling mechanism invoked should apply more universally. 

To ground our approach in the physics of black holes, we have relied
on a number of key principles and results from classical general
relativity and field theory, including the implications of the
equivalence principle \cite{Susskind93} for the field-theoretic state
at the event horizon, the nonexistence of a ``bleaching''
mechanism \cite{bleach}, and the requirement for some thermalization
or scrambling mechanism \cite{Hayden07,Sekino08,Giddings07} that
allows information from deep inside a black hole to reach the surface
before radiating away \cite{fnB} (although our results are largely
insensitive to the time scale and hence the underlying scrambling
mechanism).

Previously, the {\it no-hiding theorem} \cite{me,Kretschmann} was used to
prove that Hawking's prediction of featureless radiation implied that the
information about the in-fallen matter could not be in the radiation
field but must reside in the remainder of Hilbert space---then
presumed to be the black hole interior. That work presented a strong
form of the black hole information paradox pitting the predictions of
general relativity against those of quantum mechanics \cite{me}. Here,
we have shown that transevent horizon entanglement provides a way out,
since now the ``remainder of Hilbert space'' comprises both the black
hole interior and external neighborhood. Because the evaporating
black hole actually involves three subsystems, the information may be
encoded within them as pure correlations via a quantum one-time
pad \cite{me}, so the information remains inaccessible from any
one subsystem.

Importantly, the detailed physics of black holes (inside the event
horizon) remains beyond the scope of this Letter. Thus, this Letter
leaves mysterious those long-standing questions about the internal
dynamics of black holes that would require knowledge of the geometry
well within the black hole and extensive field-theoretic calculations
or even a theory of quantum gravity to be addressed. The very assumption
of unitarity is one such question. Another is our positing of a finite
entanglement entropy across the event horizon, without a detailed
field-theoretic description of how this should be
calculated \cite{tHooft85,Hawking01,Brustein06,Emparan06,Pad10}.
Similarly, the dynamics of the entangled degrees of
freedom exterior to the black hole remains unclear.
Finally, we assume the existence of some global thermalization process 
that leads to complete scrambling of the information encoded within the 
black hole.

The simultaneous encoding of information externally (in the combined
radiation and external neighborhood) and ``internally'' (if one slightly
{\it stretches\/} the horizon to envelop the bulk of the external
neighborhood entanglement in addition to the black hole interior)
is reminiscent of the principle of black hole complementarity
\cite{Susskind93}. This principle was introduced to account for the
apparent cloning suggested by the possibility of choosing a ``nice time''
slice through the black hole spacetime that crosses most of the
outgoing radiation as well as the collapsing body well inside the
event horizon but still far from the singularity \cite{Lowe95}.
Interpreted in the context of our work here, if such slices are drawn
after the encoding of the information into the tripartite quantum
one-time pad, the ``cloning'' would be a manifestation of the multiple
ways of reading out the information from the tripartite structure.
If such slices are drawn before the encoding occurred, then too
little of the outgoing radiation would be crossed for a potential
violation of the no-cloning theorem (note that the number of qubits
radiated may be used as a surrogate for a time coordinate).

Our results indicate that, except for the very final, vanishingly
small fraction of a (large) black hole's lifetime, the Hawking
radiation is completely uncorrelated with the state of the in-fallen
matter. Thus, the behavior Hawking found so indicative of a loss
of unitarity is in fact completely generic for unitarily evolving,
entangled black holes, requiring no specially designed evolution. 
Of course, by assuming unitarity from the outset, we cannot directly
address the black hole information paradox. Rather, our result
dissociates completely information-free radiation from a loss of
unitarity and hence undermines the very logic used to formulate the
paradox. 

Finally, in light of their curious equality, it has previously been 
conjectured that a black hole's thermodynamic entropy is actually 
entropy of entanglement \cite{tHooft85,Hawking01,Brustein06,Emparan06}.
Indeed, it unavoidably holds for some types of extremal black
holes \cite{Hawking01,Brustein06} and even 
allows their entropy to be computed at the microscopic level 
\cite{Emparan06}. The conventional riposte to this conjecture
is made by noting that the entropy of entanglement of quantum fields
piercing a black hole's event horizon would be proportional
to the number of matter fields that exist, but, since a black hole's
thermodynamic entropy is purely geometric, there should be no
{\it a priori\/} relationship between these quantities (see, e.g.,
Ref.~[\onlinecite{Nishioka09}]; for a counterargument see
Ref.~[\onlinecite{Brout}]). By studying dynamically evolving black holes,
not merely static ones \cite{tHooft85,Hawking01,Brustein06,Emparan06},
we now counter this conventional riposte. Equating a black hole's
entropy with entropy of entanglement suggests the existence of a sum
rule to constrain the number and types of matter fields in any
fundamental theory.

S.L.B.\ acknowledges the kind hospitality of the A.\ Watssman Institute
for Innovative Thinking, where this work was initiated.  
K.{\.Z}.\ acknowledges support by the Deutsche
Forschungsgemeinschaft under Project No.\ SFB/TR12 and
under the Project No.\  DEC-2011/02/A/ST1/00119,
financed by the Polish National Science Center.
%
The authors thank N.\ Cohen,
M.\ Patra, and H.-J.\ Sommers for fruitful discussions.

%
\vskip 0.1in
{\it Note added}.---Our ``energetic curtain'' (first coined in
Ref.\ \cite{SLB2009}) appears to be the same phenomenon recently
called a ``firewall'' \cite{firewall}.

\section*{SUPPLEMENTAL MATERIAL}

\section{I.\ Structure and interpretation of decoupling theorems}

Decoupling theorems effectively describe the performance of random
quantum error correction codes (QECC), whereby the quantum state
to be protected $|\psi\rangle_{\text{input}}$ is embedded into a
larger `code' Hilbert space by
$|\psi\rangle\rightarrow |\psi\rangle\otimes |\phi_0\rangle$ followed by
its encoding by a Haar-random unitary $U$ acting on the code space. Thus
\begin{equation}
\text{QECC}:|\psi\rangle_{\text{input}}\rightarrow
|\Psi\rangle_{\text{code}}=(U|\psi\rangle)_{\text{code}},
\label{qecc}
\end{equation}
where $|\Psi\rangle_{\text{code}}$ is the larger dimension code state
and in the last expression we have suppressed the ancillary subsystem
in standard state $|\phi_0\rangle$.

A subtlety we should mention is that the proof of the quantum
decoupling theorem relies on the input state, which we wish to later
locate, being entangled with a reference subsystem (ref). Thus, for
example, Eq.~(\ref{qecc}) becomes
\begin{equation}
\openone\otimes \text{QECC}:\sum_i|i\rangle_{\text{ref}}
\otimes |i\rangle_{\text{input}}
\rightarrow
\sum_i|i\rangle_{\text{ref}}\otimes(U|i\rangle)_{\text{code}}
\label{eqecc}
\end{equation}
where we have suppressed normalization for convenience.
This is a powerful step because it effectively allows us to utilize
entanglement monogamy to precisely pin down where our encoded
subsystem may be located.

With access to a sufficiently large piece of the code subspace,
decoupling theorems tell us how well we can, in principle,
reconstruct the original state (with its full entanglement to the
reference in tact). In particular, for a $k$-qubit input state encoded
into an $n$-qubit code state, there exists a threshold of $\frac{1}{2}(n+k)$
qubits above which access to more than this number of qubits of the
code state allows near ideal reconstruction of the original state.
More precisely, access to any $\frac{1}{2}(n+k)+c$ qubits from the code
state allows reconstruction \cite{Hayden07app} of the original state with
a mean fidelity of reconstruction (averaged over random encodings) bounded
below by $1-2^{-c}$. Since the `excess' unaccessed qubits of the code
are not needed for the reconstruction, they have effectively decoupled
from the original state and so the reconstruction protocol is unaffected
by any errors that occur on these excess qubits. Thus the decoupling
theorem quantifies the performance of random quantum error correcting
codes.

Importantly, the proofs of (quantum) decoupling theorems are
non-constructive. They only demonstrate the existence of a reconstructing
unitary with the claimed performance, but do not say how it may be
made.

\section{II.\ Decoupling in a classical setting}

Here we paraphrase the discussion in Ref.~[\onlinecite{Hayden07app}] for a
simple version of a decoupling result in a classical setting. Consider
a $k$-bit plaintext message randomly encoded into an $n$-bit ciphertext
string. The codebook for this code will consist of $2^k$ $n$-bit random
codewords and their associated $k$-bit messages. Obviously, anyone with
access to the codebook and any specific encoded message will be able to
exactly decode it. However, knowing the codebook allows one to do almost
as well with just a few more than $k$ bits of the encoded message 
(indeed {\it any\/} $k$ plus a few bits) of the encoded message.

As noted in Ref.~[\onlinecite{Hayden07app}], given access to only $k+c$ bits
of (and their location in) the encoded message, one can eliminate
many of the potential entries in the codebook, thus narrowing down the
possible message. To estimate the probability for this procedure
to identify any particular message from $k+c$ bits we may treat matches
as uniformly random (for our randomly generated codebook).
The probability of a random match between $k+c$ bits from a specific
encoded message and the identically located $k+c$ bits from any specific
codeword in the codebook will then be $2^{-(k+c)}$. Given therefore that
there are $2^k$ possible messages to distinguish between, the probability
of failure to identify the correct message will be $2^k 2^{-(k+c)}=2^{-c}$.
Finally, the probability with which access to any $k+c$ bits of
the encoded message (and the codebook) allows one to have successfully
reconstructed the original plaintext message (what one might call the
fidelity of reconstruction) is just $1-2^{-c}$.

We might note some significant differences between this classical
decoupling and quantum decoupling results. First, for the case where
$k\ll n$, classical decoupling allows many reconstructions of the
original message from completely distinct subsets of bits from the
encoded message --- the classical information can be cloned in this
manner. By contrast, for the analogous quantum encoding [as in
Eq.~(\ref{eqecc})] access to $\frac{1}{2}(n+k)+c$ qubits of the
encoded state are needed to achieve a reconstruction fidelity of
$1-2^{-c}$. So quantum cloning is strictly prohibited. Nonetheless, any
$\frac{1}{2}(n+k)+c$ qubits are adequate for this purpose. Second, in
the classical setting the reconstruction protocol is trivial given access
to the codebook, whereas the analogous reconstruction protocol in the
quantum case is only shown to exist (the proof is non-constructive)
but may require full knowledge of the encoding random unitary $U$,
which would be an extreme burden in any even moderately
high-dimensional scenario.

\section{III.\ Penalty for disentanglement across a hypothetical boundary}

Here we investigate the energy penalty one must pay for creating
a disentangled (i.e., separable) state across a hypothetical boundary. 
We are not here going to consider the effects of real boundary conditions
on the state of a quantum system, merely the effect of a constraint
on the state space so as to exclude entangled states across a
non-physical (fictive) boundary. Indeed, the equivalence principle has
been argued \cite{Susskind93app} to imply that freely falling observers
see nothing physical as they pass the event horizon.

Consider $M$ coupled Harmonic oscillators with Hamiltonian
\begin{equation}
H=\frac{1}{2}\sum_{i=1}^M p_i^2 + \frac{1}{2}\sum_{i,j=1}^M
K_{ij}\, x_i x_j,
\label{HOHam}
\end{equation}
where $[x_i,p_j]=i\, \delta_{ij}$ and $K$ is a real symmetric (non-negative
definite) matrix. The ground state wavefunction as a function of
$\vec x\equiv (x_1,\ldots,x_M)^T$ is
\begin{equation}
\Psi(\vec x)=\frac{(\det \sqrt{K})^{1/4}}{\pi^{M/4}}
\exp(-\vec x\cdot \sqrt{K}\cdot \vec x),
\end{equation}
with ground state energy $\frac{1}{2} \text{tr}\,( \sqrt{K})$.

Let us introduce a hypothetical boundary at index $b<M$. We assign all
oscillators with indices $i\le b$ as `inside' this fictive boundary and
all other oscillators as `outside'. It is natural to partition the
coupling matrix $K$ into blocks as
\begin{equation}
K=\left(
\begin{array}{cc}
K_{\text{in}} & Q \\
Q^T & K_{\text{out}}
\end{array}
\right).
\end{equation}
where $K_{\text{in}}$ is a $b\times b$ symmetric matrix and 
$K_{\text{out}}$ is an $(M-b)\times (M-b)$ symmetric matrix.
The Hamiltonian of Eq.~(\ref{HOHam}) may be rewritten as
\begin{eqnarray}
H&=&\frac{1}{2}\bigl(\vec p_{\text{in}}^{\,2} 
+\vec x_{\text{in}}\cdot K_{\text{in}} \cdot \vec x_{\text{in}}
+ \vec p_{\text{out}}^{\,2}
+\vec x_{\text{out}}\cdot K_{\text{out}} \cdot \vec x_{\text{out}}\bigr)
\nonumber \\
&& +\,\vec x_{\text{in}}\cdot Q\cdot \vec x_{\text{out}},
\label{QHam}
\end{eqnarray}
where $\vec x=\vec x_{\text{in}}\oplus \vec x_{\text{out}}$ decomposes
$\vec x$ into a $b$-dimensional vector $\vec x_{\text{in}}$ and
an $(M-b)$-dimensional vector $\vec x_{\text{out}}$. This effectively
decomposes the full $M$-oscillator Hilbert space ${\cal H}_{\text{total}}$
into a tensor product
${\cal H}_{\text{total}}={\cal H}_{\text{in}}\otimes {\cal H}_{\text{out}}$.

\vskip 0.1truein
\noindent
{\bf Theorem:} The general separable state across 
${\cal H}_{\text{in}}\otimes {\cal H}_{\text{out}}$ with lowest energy
for Hamiltonian~(\ref{QHam}) has energy above the ground state of
\begin{equation}
E_{\text{penalty}}\equiv\frac{1}{2}\bigl[
{\text{tr}}\,(\sqrt{K_{\text{in}}}) + {\text{tr}}\,(\sqrt{K_{\text{out}}})
-{\text{tr}}\,(\sqrt{K}) \bigr].
\label{Eest}
\end{equation}

We call this the minimal `energy penalty' for ensuring the separability
of a state across a hypothetical boundary.

\vskip 0.1truein
\noindent
{\bf Proof:} A general separable state is just the convex sum over
states of the form $\rho_{\text{in}}\otimes\rho_{\text{out}}$, where
without loss of generality we may treat $\rho_{\text{in}}$ and
$\rho_{\text{out}}$ as pure states. A lower bound to the energy
expectation of a general separable state is therefore given by the lower
bound for the energy expectation over a single such tensor product of
pure states.

Consider now a general product of pure states. We may always write its
wavefunction as a displaced product
\begin{eqnarray}
\!\!\!\!\!&&\!\Psi_{\text{prod}}(\vec x_{\text{in}}, \vec x_{\text{out}})
\label{Gstate}\equiv \\
\!\!\!\!\!&&\!
D_{\text{in}}(\vec x_{\text{in}}^{\,0}+i\vec p_{\text{in}}^{\;0})
\Psi_{\text{in}}(\vec x_{\text{in}})
D_{\text{out}}(\vec x_{\text{out}}^{\,0}+i\vec p_{\text{out}}^{\;0})
\Psi_{\text{out}}(\vec x_{\text{out}}),\nonumber
\end{eqnarray}
where $\Psi_{\text{zero}}\equiv \Psi_{\text{in}} \Psi_{\text{out}}$
is taken to have zero mean positions and momenta. The expectation 
$\langle H\rangle_{\text{prod}}$ of Hamiltonian~(\ref{QHam}) with
respect to the general state of Eq.~(\ref{Gstate}) may now be rewritten
as an expectation over this `zero mean' state $\Psi_{\text{zero}}$ as
\begin{eqnarray}
\!\!\!&&\frac{1}{2}\Bigl[\bigl\langle\vec p_{\text{in}}^{\,2}
+\vec x_{\text{in}}\cdot K_{\text{in}} \cdot \vec x_{\text{in}}
+ \vec p_{\text{out}}^{\,2}
+\vec x_{\text{out}}\cdot K_{\text{out}} \cdot \vec x_{\text{out}}
\bigr\rangle_{\text{zero}} \nonumber \\
\!\!\!&& +\,\vec p_{\text{in}}^{\;0\,2}
+\vec x_{\text{in}}^{\,0}\cdot K_{\text{in}} \cdot \vec x_{\text{in}}^{\,0}
+ \vec p_{\text{out}}^{\;0\,2}
+\vec x_{\text{out}}^{\,0}\cdot K_{\text{out}} \cdot \vec x_{\text{out}}^{\,0}
\nonumber \\
\!\!\!&&+\,2\,\vec x_{\text{in}}^{\,0}\cdot Q\cdot \vec x_{\text{out}}^{\,0}
\Bigr].
\end{eqnarray}
Since $K$ is non-negative definite, for any vector
$\vec x^{\,0}=\vec x_{\text{in}}^{\,0}\oplus \vec x_{\text{out}}^{\,0}$
we have $\vec x^{\,0}\cdot K \cdot \vec x^{\,0}\ge 0$. Thus
\begin{eqnarray}
\!\!\!\!\!&&\langle H\rangle_{\text{prod}}\ge\\
\!\!\!\!\!&& \frac{1}{2}
\bigl\langle\vec p_{\text{in}}^{\,2}
+\vec x_{\text{in}}\cdot K_{\text{in}} \cdot \vec x_{\text{in}}
+ \vec p_{\text{out}}^{\,2}
+\vec x_{\text{out}}\cdot K_{\text{out}} \cdot \vec x_{\text{out}}
\bigr\rangle_{\text{zero}}.\nonumber
\end{eqnarray}
Note that the right-hand-side is just the expectation of the sum of a pair
of independent oscillators with individual ground state energies 
$\frac{1}{2} \text{tr}\,( \sqrt{K_{\text{in}}})$ and 
$\frac{1}{2} \text{tr}\,( \sqrt{K_{\text{out}}})$ respectively. Thus,
\begin{equation}
\langle H\rangle_{\text{prod}}\ge 
\frac{1}{2}\bigl[ \text{tr}\,( \sqrt{K_{\text{in}}})
+\text{tr}\,( \sqrt{K_{\text{out}}})\bigr].
\end{equation}
Further, since these independent product ground states have zero means,
this lower bound is achieved.
${~}$\hfill \rule{2mm}{2mm}\\

In order to see how this separability penalty appears in a field
theoretic setting consider a free scalar field with Hamiltonian
\begin{equation}
H=\frac{1}{2}\int d^3 x\,
\bigl[\pi^2+(\vec \nabla \varphi)^2\bigr],
\end{equation}
where $\pi=\partial_t \varphi$ is the conjugate momentum for the quantum
field $\varphi$ and these satisfy the equal-time canonical commutation
relations
\begin{equation}
\bigl[\varphi(t,\vec x),\pi(t,\vec x')\bigr]=i\, \delta(\vec x-\vec x').
\end{equation}

Following Srednicki \cite{Srednickiapp}, we introduce a lattice of discrete
points with equal spacing $a$ in the radial direction. Furthermore, the
field is placed in a spherical box of radius $A=(N+1)a$ and the field
is taken to vanish at the (real) boundary at $A$. The field and its
conjugate momentum can be decomposed into partial waves $\varphi_{j,lm}$
and $\pi_{j,lm}$ satisfying the equal time commutation relation
\begin{equation}
[\varphi_{j,lm}, \pi_{j',l'm'}]=i\,\delta_{jj'} \delta_{ll'} \delta_{mm'},
\end{equation}
where $j a$ gives the discrete radial coordinate and $\{l,m\}$
label the partial waves' angular momentum. The discretized
Hamiltonian then becomes $H=\sum_{l,m} H_{lm}$, with \cite{Srednickiapp}
\begin{eqnarray}
H_{lm}&=&\frac{1}{2a}\sum_{j=1}^N\left[
\pi_{j,lm}^2 +(j+\frac{1}{2})^2\Bigl(\frac{\varphi_{j,lm}}{j}
-\frac{\varphi_{j+1,lm}}{j+1}\Bigr)^2 \right.\nonumber\\
&&\phantom{\frac{1}{2a}\sum_{j=1}^N~}\left.
+\,\frac{l(l+1)}{j^2}\varphi_{j,lm}^2\right].
\label{Hfield}
\end{eqnarray}

Numerical calculations of $E_{\text{penalty}}$ from Eq.~(\ref{Eest}) for
this discretized Hamiltonian yield
\begin{equation}
E_{\text{penalty}}\simeq 0.05\, \frac{(N+1)^2}{a}=
0.05\, \frac{A^2}{a^3},
\end{equation}
where the hypothetical boundary index is chosen as $b=N/2$ (across a
range of even $N$ from $50$ to $100$). This penalty diverges as the
cube of the ultraviolet regulator $1/a$. Thus we expect pure quantum
states where entanglement has essentially vanished across a hypothetical
boundary to have very large energies.

\section{IV.\ Uniform entanglement}

Because one of the key claims in the paper is about loss of trans-event
horizon entanglement, we shall repeat the key calculation here
for a black hole with trans-event horizon entanglement, but where,
for simplicity, that entanglement is taken to be uniform. This
allows us to repeat the analysis solely using results already available in
the literature.

Consider black hole evaporation with uniform trans-event
horizon entanglement as
\begin{equation}
\frac{1}{\sqrt{E}}\sum_{j=1}^{E}
|j\rangle_{\text{int}}\otimes|j\rangle_{\text{ext}}\rightarrow
\frac{1}{\sqrt{E}}\sum_{j=1}^{E}
(U|j\rangle)_{\text{BR}} \otimes|j\rangle_{\text{ext}}.
\end{equation}
Here $\log_2 E$ is the entropy of entanglement between the external (ext)
neighborhood and the interior of the black hole. Except for the
interpretation of the source of entanglement, this model has been
recently analyzed by Ref.~[\onlinecite{Hayden07app}]. We may therefore quote
their key result in our terms: For any positive $c$, once
$\frac{1}{2} S_{\text{BH}}+\frac{1}{2} \log_2 E + c$ qubits have radiated
away [this is just the $\frac{1}{2}(n+k)+c$ qubits required as discussed
in the first section of this Supplementary Material], the 
trans-event horizon entanglement between the external neighborhood and
the interior subsystems will have virtually vanished, with it appearing
instead (with a fidelity of at least $1-2^{-c}$) as entanglement between
the external neighborhood and the outgoing radiation. Here (as in our
manuscript) $c$ is a free parameter, but will be dwarfed by any of the
entropies involved.

Repeating the argument from our manuscript, this loss must be delayed
until the black hole has evaporated to roughly the Planck scale.
(Indeed, section III of this Supplementary Material provides
energy estimates for the departure from vacuum across the event horizon
when trans-event horizon entanglement is lost.) Such a delay implies that
roughly $S_{\text{BH}}$ qubits must have already been radiated before
such loss occurs, in which case
\begin{equation}
\log_2 E\approx S_{\text{BH}}.
\end{equation}

In section V below, we shall see that when the uniform entanglement of
the above analysis is replaced with general trans-event horizon
entanglement, the measure of entanglement $\log_2 E$ is replaced by
the R\'enyi entropy $H^{(1/2)}(\rho_{\text{ext}})$. This replacement
is unchanged in the presence of in-fallen matter (also section VI).

\section{V.\ Formalism for General entanglement}
\label{decoupling}

Note that all R\'enyi entropies are bounded above by the logarithm of
the Hilbert space dimension, so
 $0\le H^{(q)}(\rho_{\text{ext}})\le n\equiv S_{\text{BH}}$
for the state we study. Of particular interest to us here will be two
R\'enyi entropies for $q=\frac{1}{2}$, $2$, so
\begin{eqnarray}
H^{(1/2)}(\rho_{\text{ext}}) &=&\log_2
 \,\bigl[({\text{tr}}\;\sqrt{\rho_{\text{ext}}}\,)^2\bigr] \nonumber \\
H^{(2)}(\rho_{\text{ext}}) &=&-\log_2
\,({\text{tr}}\;{\rho^2_{\text{ext}}}\,).
\end{eqnarray}
(In the limit of $q\rightarrow 1$ the R\'enyi entropy reduces to the
more familiar von Neumann entropy.)

Our key result is based on our generalization (theorem below) of the
decoupling theorem of Ref.~[\onlinecite{Abey06app}]. Consider now the
tripartite state
\begin{equation}
\rho_{XYZ}^{\text{~}}=
\rho_{XY_1Y_2Z}^{\text{~}},
\end{equation}
where the joint subsystems $Y=Y_1Y_2$ will be decomposed as either
the radiation subsystem and interior black hole subsystem $RB$ or vice-versa
$BR$. This allows us to define
\begin{equation}
\sigma_{XY_2Z}^U\equiv
\text{tr}_{Y_1}^{\text{~}} \bigl(U_{Y}\;
\rho_{XYZ}^{\text{~}}\; U_{Y}^\dagger\bigr).
\end{equation}
In keeping with the naming convention of Ref.~[\onlinecite{Abey06app}],
we call the result below the Mother-in-law decoupling theorem.


\vskip 0.1truein
\noindent
{\bf Generalized decoupling theorem:}
\begin{eqnarray}
&&\biggl(\int_{U\in U(Y)}dU\,\bigl\| \sigma_{XY_2Z}^U-
\sigma_{X}^U\otimes \sigma_{Y_2Z}^U\bigr\|_1\biggr)^2\nonumber\\
&\le& {\text{tr}}\; \rho_{X}^{2\nu}\; {\text{tr}}\; \rho_{Z}^{2\mu}
\biggl\{\Bigl[ {\text{tr}}\; \rho_{XZ}^{2}
        ( \rho_{X}^{-2\nu}\otimes \rho_{Z}^{-2\mu})\nonumber\\
&&\phantom{{\text{tr}}\; \rho_{X}^{2\nu}\; {\text{tr}}\; \rho_{Z}^{2\mu}}
\;-2\, {\text{tr}}\; \rho_{XZ}^{\text{~}}
   (\rho_{X}^{1-2\nu}\otimes \rho_{Z}^{1-2\mu})\nonumber \\
&&\phantom{{\text{tr}}\; \rho_{X}^{2\nu}\; {\text{tr}}\; \rho_{Z}^{2\mu}}\;
+{\text{tr}}\; \rho_{X}^{2-2\nu}\; {\text{tr}}\; \rho_{Z}^{2-2\mu}\Bigr]
\nonumber\\
&&\phantom{{\text{tr}}\; \rho_{X}^{2\nu}\; {\text{tr}} }
+\frac{Y_2}{Y_1}\Bigl[
{\text{tr}}\; \rho_{XYZ}^{2}
        ( \rho_{X}^{-2\nu}\otimes \rho_{Z}^{-2\mu}) \nonumber \\
&&\phantom{{\text{tr}}\; \rho_{X}^{2\nu}\; {\text{tr}}~~~~~~~~}
+{\text{tr}}\; \rho_{X}^{2-2\nu}\;
{\text{tr}}\; \rho_{YZ}^{2}\;\rho_{Z}^{-2\mu}\Bigr]
\biggr\} \label{step1}\\
&\le & \frac{Y_2}{Y_1}\, {\text{tr}}\; \rho_{X}^{2\nu}\;
{\text{tr}}\; \rho_{Z}^{2\mu}\,
\Bigl[
{\text{tr}}\; \rho_{XYZ}^{2}
        ( \rho_{X}^{-2\nu}\otimes \rho_{Z}^{-2\mu}) \nonumber \\
&&\phantom{\frac{Y_2}{Y_1}\, {\text{tr}}\; \rho_{X}^{2\nu}\; 
{\text{tr}}\; \rho_{Z}^{2\mu}\Bigl[\,}
+{\text{tr}}\; \rho_{X}^{2-2\nu}\; 
{\text{tr}}\; \rho_{YZ}^{2}\;\rho_{Z}^{-2\mu}\Bigr] \label{step2} \\
&\le & 2\, \frac{Y_2}{Y_1}\,
2^{H_X^{\text{~}}+H_Z^{\text{~}}},
\label{step3}
\end{eqnarray}
where $H_A^{\text{~}}\equiv H^{(1/2)}(\rho_A^{\text{~}})$,
$0\le 2\nu,2\mu\le 1$, and the trace norm is defined by
$\|X\|_1\equiv \text{tr}\, |X|$. Recall from our manuscript, that we
reuse subsystem labels for Hilbert space dimensionalities, thus
here $Y_2/Y_1$ denotes the ratio of their Hilbert space dimensions.
Note that here, to go from Eq.~(\ref{step1}) to Eq.~(\ref{step2}),
we would require
$\rho_{XZ}^{\text{~}}=\rho_X^{\text{~}}\otimes \rho_Z^{\text{~}}$; and
to go from Eq.~(\ref{step2}) to Eq.~(\ref{step3}), we would require
$\rho_{XYZ}^{\text{~}}$ is pure and we take $2\nu=2\mu=\frac{1}{2}$.
\\

\noindent
{\bf Proof:}
Using the Cauchy-Schwarz inequality we may write
\begin{eqnarray}
&&\bigl\| \sigma_{XY_2Z}^U-
\sigma_{X}^U\otimes \sigma_{Y_2Z}^U\bigr\|_1 \\
&\le& \bigl\| \rho_{X}^{\nu}\otimes\openone_{Y_2}^{\text{~}}\otimes
\rho_{Z}^{\mu} \bigr\|_2\nonumber\\
&&\times
\bigl\| \rho_{X}^{-\nu}\otimes \rho_{Z}^{-\mu}(\sigma_{XY_2Z}^U-
\sigma_{X}^U\otimes \sigma_{Y_2Z}^U)\bigr\|_2, \nonumber
\end{eqnarray}
where without loss of generality we may assume that $\rho_{X}^{\nu}$
and $\rho_{Z}^{\mu}$ are invertible; then using the methods already
outlined in Ref.~[\onlinecite{Abey06app}] the results are easily obtained.
{$~$}\hfill \rule{2mm}{2mm}\\
We note that the statement of the result reduces to the conventional
decoupling theorem for the choice $\nu=0$ and subsystem $Z$ is
one-dimensional.

Of particular interest here is the case where $2\nu=\frac{1}{2}$ and
$\rho_{\text{ext},Y}$ is pure, which gives
\begin{equation}
\int_{U\in U(Y)}dU\,\bigl\| \sigma_{\text{ext},Y_2}^U-
\sigma_{\text{ext}}^U\otimes \sigma_{Y_2}^U\bigr\|_1
\le \Bigl(2\,\frac{Y_2}{Y_1}\,2^{H_{\text{ext}}}\Bigr)^{\frac{1}{2}},
\end{equation}
with $H_{\text{ext}}\equiv H^{(1/2)}(\rho_{\text{ext}})$.

Now $1-F(\rho,\sigma) \le \frac{1}{2}\|\rho-\sigma\|_1$, where the fidelity
is defined by $F(\rho,\sigma)\equiv \|\sqrt{\rho}\sqrt{\sigma}\|_1$.
As a consequence, the fidelity with which the initial trans-event
horizon entanglement is encoded within the combined $\text{ext},Y_1$
subsystem is bounded below by $1-\sqrt{2^{H_{\text{ext}}} Y_2/Y_1}$.
Now allowing this in turn to be bounded from below by $1-2^{-c}$ and
choosing $Y_1=R$ and $Y_2=B$ and recalling that $BR=2^{S_{\text{BH}}}$
gives the result quoted in our manuscript [Eq.~(3) there].

Interestingly, the opposite choice $Y_1=B$ and $Y_2=R$ tells us that,
for any positive $c$, for fewer than
$\frac{1}{2}[S_{\text{BH}}-H^{(1/2)}(\rho_{\text{ext}})]-c$
qubits radiated away, the initial trans-event horizon entanglement
remains encoded between the external neighborhood and the interior
subsystems with fidelity of at least $1-2^{-c}$. This effectively
gives the number of qubits that must be radiated before trans-event
horizon entanglement {\it begins\/} to be reduced from its initial
value. Of particular interest is the case when
$H^{(1/2)}(\rho_{\text{ext}})\approx S_{\text{BH}}$ for which we would
conclude that the trans-event horizon entanglement begins to be
depleted by radiation almost immediately.

\section{VI.\ Entanglement loss in the presence of in-fallen matter}

The unitary evaporation of an entangled black hole in the
presence of in-fallen matter was argued in our manuscript to be 
described by
\begin{eqnarray}
&&\frac{1}{\sqrt{K}}\sum_{i=1}^K |i\rangle_{\text{ref}}\otimes
\sum_{j}\sqrt{p_j}\,(|i\rangle\otimes |j\rangle\oplus 0)_{\text{int}}
\otimes|j\rangle_{\text{ext}}\label{entUmodelapp} \\
&\rightarrow&
\frac{1}{\sqrt{K}}\sum_{i=1}^K |i\rangle_{\text{ref}}\otimes
\!\sum_{j}\sqrt{p_j}\,[U(|i\rangle\otimes |j\rangle\oplus 0)]_{BR}
\otimes|j\rangle_{\text{ext}}, \nonumber
\end{eqnarray}
where $\log_2 K\equiv S_{\text{matter}}$ is the number of qubits of
quantum information in the in-fallen matter.

It is now straightforward to apply the generalized decoupling
theorem above to show that, for an arbitrary positive number $c$, when 
the number of qubits radiated reaches
\begin{equation}
\log_2 R = S_{\text{BH}}-\frac{1}{2}\chi^{(1/2)} +c,
\end{equation}
then the trans-event horizon entanglement has effectively vanished
and instead has been transferred to entanglement between the
external neighborhood and the outgoing radiation, with
a fidelity of at least $1-2^{-c}$. Recall from our manuscript that
the number of unentangled qubits initially within the black hole
is roughly quantified by
\begin{equation}
\chi^{(q)}\equiv S_{\text{BH}} - S_{\text{matter}}
 -H^{(q)}(\rho_{\text{ext}})\ge 0, \label{chiqapp}
\end{equation}
with $q$ of order unity.

Repeating the argument from our manuscript, unless this occurs
when $\log_2 R \approx S_{\text{BH}}$ then a noticeable violation of
the equivalence principle will occur. This implies that
\begin{equation}
\chi^{(1/2)}\lll S_{\text{BH}},
\end{equation}
or equivalently, that
\begin{equation}
S_{\text{BH}}\approx H^{(1/2)}(\rho_{\text{ext}})+S_{\text{matter}}
\approx H^{(1/2)}(\rho_{\text{ext}}),
\end{equation}
since from 't Hooft's bound, the entropic content of matter
is only a vanishingly small fraction of the thermodynamic entropy
of the black hole, i.e., $S_{\text{matter}}\lll S_{\text{BH}}$.

\section{VII.\ Where's the Planck scale?}

In the first part of the manuscript, we show that a black hole's
thermodynamic entropy must be very well approximated by its
entropy of entanglement across the event horizon. The proof relied
on preservation of the equivalence principle prior to the black hole
having evaporated to the Planck scale. However, what defines the
beginning of the Planck scale for black holes? 

A universal feature of black holes is their thermodynamic entropy or
(essentially up to a constant prefactor) their surface area. We shall
therefore use the entropy (in bits) as a measure of size of a black hole.
As we wish to avoid making claims about the physics of Planck scale
black holes, we shall suppose there is some size, above which Planck
scale effects are negligible (in particular, above which the predictions
of the equivalence principle are left in tact). Stated conversely,
we shall suppose that entry into the Planck scale regime, where effects
on the equivalence principle begin to become non-negligible, occurs at
some generic size (or equivalently entropy). In particular, we take this
entry into the Planck scale for black holes of entropy smaller than
\begin{equation}
S_{\text{BH}}^{\text{Planckian}} \lesssim 2^p.
\label{entry}
\end{equation}
It will turn out that virtually any choice for $p$ makes no difference
to our analysis since the entry-point entropy so defined will be
dwarfed by those of that of typical large black holes (e.g.,
a stellar mass black hole has thermodynamic entropy of ${10^{10}}^{77}$).

To see how the argument runs, we must determine the thermodynamic entropy
of a black hole evaporating according to Eq.~(\ref{entUmodelapp}). We shall
suppose the von Neumann entropy computed from Eq.~(\ref{entUmodelapp}) is
a good estimate for the thermodynamic entropy. Evolution corresponds to
the radiation subsystem $R$ becoming an ever larger portion of the
initial black hole Hilbert space and the remaining black hole interior 
subsystem $B$ becoming an ever shrinking portion, subject to the 
constraint that
\begin{equation}
2^{S_{\text{BH}}}=BR,
\end{equation}
where one should recall that we reuse subsystem labels as their 
corresponding Hilbert space dimensionalities. During evaporation, a
simple upper bound to the von Neumann entropy of the black hole interior
$S(B)$ is given by the logarithm of its dimensionality. Hence
\begin{equation}
S(B)\leq S_{\text{BH}} -\log_2 R.
\end{equation}

For a lower bound we can use the negative logarithm of the so-called purity
\begin{equation}
S(B)\geq -\log_2 \langle\!\langle 
{\text{tr}}\; (\rho_B^U)^2 \rangle\!\rangle_U,
\end{equation}
where $\rho_B^U$ is the reduced substate on the black hole interior
of Eq.~(\ref{entUmodelapp}), and
$\langle\!\langle \cdots \rangle\!\rangle_U$ denotes averaging over
the random unitary $U$ with the Haar measure. Using standard methods 
\cite{Abey06app}, the purity can be easily estimated in the latter stages
of evaporation to be
\begin{equation}
\langle\!\langle
{\text{tr}}\; (\rho_B^U)^2 \rangle\!\rangle_U
\simeq \frac{B(R^2-1)}{(BR)^2-1} \simeq \frac{R}{2^S_{\text{BH}}},
\end{equation}
and hence $S(B)\gtrsim S_{\text{BH}} -\log_2 R$. These bounds
imply that during the latter stages of evaporation via
Eq.~(\ref{entUmodelapp}), a black hole's von Neumann entropy will be
\begin{equation}
S(B)\simeq S_{\text{BH}} -\log_2 R.
\label{SB}
\end{equation}

Combining Eqs.~(\ref{entry}) and~(\ref{SB}) we see that for a large
black hole to have evaporated to just above the Planck scale 
(prior to any need to invoke Planck-scale physics), it must have emitted
virtually all its initial entropy as Hawking radiation. In other words,
as one approaches the Planck scale, one has to very high precision
that
\begin{equation}
\log_2 R \approx S_{\text{BH}}.
\end{equation}

\section{VIII.\ Information retrieval from pure-state black holes}

As noted in our manuscript, the description of an evaporating
black hole via
\begin{equation}
|i\rangle_{\text{int}}\rightarrow (U|i\rangle)_{\text{RB}}. \label{Umodelapp}
\end{equation}
is not new. This was originally formulated \cite{Page93app}
assuming that all the in-falling matter (and the black-hole itself)
was in a pure state $|i\rangle$. In other words, it is assumed that
initially there is no trans-event horizon entanglement. It should be
noted, that prior to Ref.~[\onlinecite{B10app}] this evolution was not
connected to or claimed to be supported by any microscopic mechanism.

The original analysis suggested that a `discernible information'
(corresponding to the deficit of the entropy of a subsystem from
its maximal value) would yield a suitable metric for information
content in the radiation \cite{Page93app}. In order to find the
``typical'' behavior of an evaporating black hole it calculated
the mean discernible information averaged over random
unitaries \cite{Page93app}.

Starting with a pure-state interior, the mean discernible information
of the radiation remains almost zero until half the qubits of the
initial black hole had been radiated, after which it rises at
the rate of roughly two bits for every qubit radiated \cite{Page93app}.
This behavior suggests that first entanglement is created, followed by
dense coding \cite{Bennett92app} of {\it classical\/} information about the
initial state.

In order to get a much clearer picture of quantum information flow
in Eq.~(\ref{Umodelapp}) we can rely on the decoupling theorem.
In particular, entangling the state of the in-fallen matter with some
distant reference (ref) subsystem, allows one to track the flow of
quantum information \cite{meapp,Hayden07app}. In this way Eq.~(\ref{Umodelapp})
becomes
\begin{equation}
\frac{1}{\sqrt{K}}\sum_{i=1}^{K}
|i\rangle_{\text{ref}}\otimes|i\rangle_{\text{int}}\rightarrow
\frac{1}{\sqrt{K}}\sum_{i=1}^{K}
|i\rangle_{\text{ref}}\otimes(U|i\rangle)_{\text{BR}},
\label{HaydenPreskill}
\end{equation}
recall that $\log_2 K\equiv S_{\text{matter}}$ is the number of qubits
describing the quantum state of the matter used to form the otherwise
pure-state black hole. Using the decoupling theorem \cite{Abey06app}
we may show that, for any positive number $c$, prior to
$\frac{1}{2}(S_{\text{BH}}-S_{\text{matter}})-c$ qubits having been
radiated, the quantum information about the in-fallen matter
is encoded within the black hole interior with fidelity at least
$1-2^{-c}$; whereas after a further $S_{\text{matter}} + 2c'$ qubits
have been radiated, for arbitrary positive $c'$, the information
about the in-fallen matter is encoded within the radiation with
fidelity at least $1-2^{-c'}$ (see also Ref.~[\onlinecite{Hayden07app}]
for this latter result). The quantum information about the in-fallen
matter naively appears to leave in a narrow `pulse' at the radiation
emission rate; this pulse occurs just as half of the black hole's
qubits have radiated away.

Now consider what happens if additional matter is dumped into the 
black hole after its creation. Following Ref.~[\onlinecite{Hayden07app}], 
we model this process via cascaded random unitaries on the black hole 
interior --- one unitary before each radiated qubit. (Naturally, any
such analysis relies on a very short global thermalization time for
the black hole. An assumption which was not needed for any of the
results quoted in our manuscript itself.) Within the pure-state black
hole of Eq.~(\ref{HaydenPreskill}), it was argued \cite{Hayden07app} 
that {\it after\/} half of the initial qubits had radiated away, any
information about matter subsequently falling into the black hole
would be ``reflected'' immediately at roughly the radiation emission
rate \cite{Hayden07app}. By contrast, in the early stages of evaporation
information about matter subsequently thrown in would only begin
to emerge after half of the initial qubits of the black hole had
radiated away \cite{Hayden07app}. These very different {\it behaviors\/}
in the first and second halves of its life suggest that such a black
hole acts almost as two different species: as storage during the first
half of its radiated qubits and as a reflector during the second half.

A subtle flaw to this argument of Ref.~[\onlinecite{Hayden07app}] is due
to the omission of the fact that a black hole's entropy is
non-extensive, e.g., scaling as the square of the black hole's
mass $M^2$ for the Schwarzshild family of black holes: for every $k$
qubits dumped into such a black hole, the entropy typically increases
by $O(k M)\gg k$. Likewise, the number of unentangled qubits within
the (initially pure-state) black hole will increase by $O(k M)$.
Therefore, within the cascaded unitary pure-state black hole, the
reflection described in Ref.~[\onlinecite{Hayden07app}] would not
begin immediately, but only after a large delay in time of
$O(k M^2)$. Notwithstanding the delay, the pure-state black hole
behaves effectively as two distinct species as described above.

Because this behavior seems so bizarre it is worth going back over the
key assumptions that went into it: i) that the behavior of a specific
unitary in Eq.~(\ref{Umodelapp}) is well described by Haar averages over 
all random unities; ii) that the number of qubits comprising the initial
black hole Hilbert space is $n\simeq S_{\text{BH}}$. (These two
assumptions are discussed in some detail in our manuscript.) Finally,
iii) that the black hole is {\it initially\/} in a pure state up
to a negligible amount of entanglement that may come from the matter
content. In fact, it is this last assumption which is weakest and at
odds with the well known quantum physics of condensed matter systems
and rigorous results from axiomatic field theory as discussed in
our manuscript.

\def\badstory{
\section{IX.\ Evaporation epochs}

Throughout almost the entire manuscript we use the number of
qubits radiated as Hawking radiation as a surrogate for time. In
fact, the Hawking radiation does not proceed at a uniform rate, but is
initially slow and speeds up towards the end. Since information 
about the in-fallen matter is only accessible from a highly entangled 
black hole at very late stages in the black hole evaporation let us
consider how the non-uniform evaporation of qubits appears as a function
of time. For simplicity we consider evaporation rates for Schwarzschild
black holes, but the results are unchanged for differing black hole
types for the late stages of evolution (even though that is exactly
where the evaporation is most non-uniform).

As we are only interested in scaling behavior we ignore all constant
prefactors. Recall that the area of a Schwarzschild black hole
of mass $M$ scales as $M^2$, and that its temperature scales as
$M^{-1}$. As a thermal body, therefore, a Schwarzschild black hole
radiates with a power scaling as $M^{-2}$. This means that at time
$t$ during a Schwarzschild black hole's lifetime its mass satisfies
\begin{equation}
M(t)^3 = M^3(1 -t/T),
\end{equation}
where $T$ is the black hole lifetime, which scales as $M^3$. By the
entropy-area law, the number of qubits $n$ that one may associate with
the interior of a Schwarzschild black hole of mass $M$ scales as
$n\propto M^2$. In terms of qubits, therefore, the time evolution
may be written
\begin{equation}
n(t)^{3/2} = n^{3/2}(1 -t/T).
\end{equation}

Suppose some epoch of interest occurs at the final fraction $f$ of
the black hole's lifetime, i.e., at time $t=(1-f)T$.
}

\section{IX.\ Information retrieval from an entangled black hole}

Here we give explicit statements of results from decoupling summarized
in our manuscript.

Applying the decoupling theorem \cite{Abey06app} to the entangled-state
black hole of Eq.~(\ref{entUmodelapp}) allows us to show that, for any
positive number $c$, for all but the final
$S_{\text{matter}}+\frac{1}{2}\chi^{(2)}+c$ qubits radiated, the
information about the in-fallen matter is encoded in the combined space
of the external neighborhood and black hole interior with fidelity
at least $1-2^{-c}$. Similarly, for any positive $c'$, for all
but the initial $S_{\text{matter}}+\frac{1}{2}\chi^{(2)}+c'$ qubits
radiated, this information is encoded in the combined radiation and
external neighborhood subsystems with fidelity at least $1-2^{-c'}$.
In addition, at all times this information is encoded with unit
fidelity within the joint radiation and interior subsystems.

In other words, between the initial and final roughly
$S_{\text{matter}}+\frac{1}{2}\chi^{(2)}$ qubits radiated, the
information about the in-fallen matter is effectively {\it deleted\/}
from each individual subsystem \cite{meapp,Kretschmannapp}, instead
being encoded in any two of the three of subsystems (consisting of
the out-going radiation, the external neighborhood, and the black
hole interior). During this time, the information about the
in-fallen matter is to an excellent approximation encoded within
the perfect correlations of a {\it quantum one-time pad}
\cite{Leung02app,meapp} of these three subsystems.

Furthermore, using our generalized decoupling theorem we may show that,
for any positive $c''$, that prior to the first
$\frac{1}{2}\chi^{(1/2)} -c''$ qubits radiated,
the information about the in-fallen matter is still encoded solely
within the black hole interior, with a fidelity of at least
$1-2^{-c''}$. Similarly, for any positive $c'''$, within the final
$\frac{1}{2}\chi^{(1/2)} -c'''$ qubits radiated, the information about
the in-fallen matter is encoded within the out-going radiation,
with a fidelity of at least $1-2^{-c'''}$. Combining these with the
above results we see that both the encoding and decoding of the
tripartite quantum one-time pad occur during the radiation of roughly
$S_{\text{matter}}+\frac{1}{2}(\chi^{(2)}-\chi^{(1/2)})
\simeq S_{\text{matter}}$ qubits, i.e., the black hole's quantum
one-time pad encoding (and decoding) occurs at roughly the radiation
emission rate.

How does this entangled-state description of black hole evaporation
respond to matter subsequently swallowed after its formation? Instead
of the two distinct behaviors of storage and reflection found in the
pure-state black hole, here, any additional qubits thrown in will
immediately begin to be encoded into the tripartite one-time pad.
The decoding into the radiation subsystem of the information about
{\it all\/} the in-fallen matter will only occur at the very end of
the evaporation. (The non-extensive increase in black 
hole entropy is taken up as entanglement with the external neighborhood 
so no further delays occur.) Thus, instead of behaving almost
as two distinct species, a highly entangled-state black hole has one
principle behavior --- forming a tripartite quantum one-time pad
between the black hole interior, the external neighborhood
and the radiation from the black hole, with release of that information
only at the end of the evaporation.

Can we reconcile the information retrieval behavior of the pure-state
black hole with its entangled counterpart? Naively, if the pure-state
black hole analysis were run on twice as many qubits, but stopped
just after the information about the in-fallen matter had escaped
as a narrow pulse then there would be broad agreement between the
two types of black hole. This doubling of the number of qubits would make some
crude sense if we supposed that the pure-state black hole was not making
a split between interior and exterior at the event horizon, but
somewhat further out at some arbitrary boundary where trans-boundary
entanglement would not be participating in the evaporation. The
dimensionality of the Hilbert space within this extended boundary
would then be dominated by the product of the dimensionality of the
original black hole interior, and the nearby external neighborhood entangled
with them. This would be roughly twice the number of qubits within the
black hole interior itself. Once the original number of qubits had
evaporated away (now half the total for our extended boundary
pure-state black hole) the black hole interior would be exhausted of
Hilbert space and evaporation would cease. This suggests that
despite the general incompatibility between the two types of black hole, a
pure-state analysis, if thoughtfully set up, could capture important
features of information retrieval from an entangled-state black hole.

\section{X.\ Heuristic flow via correlations}
\label{heuristic}

The rigorous results from our manuscript may be heuristically visualized
by following how the correlations with the distant reference system
behave. For a pure tripartite state $XYZ$, these correlations satisfy
\begin{equation}
C(X\!:\!Y)+C(X\!:\!Z) = S(X), \label{monogamy}
\end{equation}
Here $S(X)$ is the von Neumann entropy for subsystem $X$ and
$C(X\!:\!Y)\equiv\frac{1}{2}[S(X)+S(Y)-S(X,Y)]$, one-half the quantum
mutual information, is a measure of correlations between subsystems
$X$ and $Y$. Relation~(\ref{monogamy}) is additive for a pure
tripartite state, so the correlations with subsystem $X$ smoothly
move from subsystems $Y$ to $Z$ and vice-versa.

\begin{figure}[ht]
\centering
\begin{tabular}{cc}
(a)$~~~~~~~~~~~~~~~~~~~~~~~~~~~~~~~~~$ &
(b)$~~~~~~~~~~~~~~~~~~~~~~~~~~~~~~~~~$ \\
  \begin{psfrags}
    \psfrag{qubitsradiated}[l]{$\scriptstyle \text{qubits radiated}$}
    \psfrag{XKB}[c]{$\scriptstyle ~~~~~~C(\text{ref}:B)$}
    \includegraphics[scale=0.45]{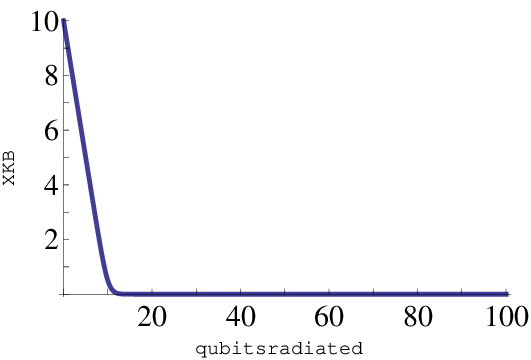}
  \end{psfrags} &
  \begin{psfrags}
    \psfrag{qubitsradiated}[l]{$\scriptstyle \text{qubits radiated}$}
    \psfrag{KAN}[c]{$\scriptstyle ~~~~~~C(\text{ref}:R,\,\text{ext})$}
    \includegraphics[scale=0.45]{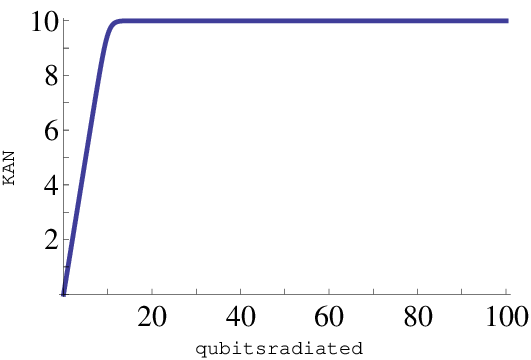}
  \end{psfrags} \\
(c)$~~~~~~~~~~~~~~~~~~~~~~~~~~~~~~~~~$ &
(d)$~~~~~~~~~~~~~~~~~~~~~~~~~~~~~~~~~$ \\
  \begin{psfrags}
    \psfrag{qubitsradiated}[l]{$\scriptstyle \text{qubits radiated}$}
    \psfrag{KBN}[c]{$\scriptstyle ~~~~~~C(\text{ref}:B,\,\text{ext})$}
    \includegraphics[scale=0.45]{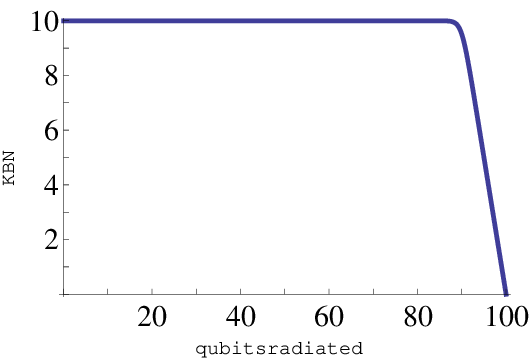}
  \end{psfrags} &
  \begin{psfrags}
    \psfrag{qubitsradiated}[l]{$\scriptstyle \text{qubits radiated}$}
    \psfrag{XKA}[c]{$\scriptstyle ~~~~~~C(\text{ref}:R)$}
    \includegraphics[scale=0.45]{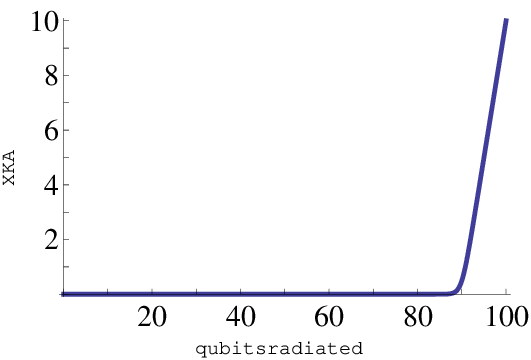} 
  \end{psfrags}
\end{tabular}
\caption{Correlations to the reference subsystem as a function of the
number of qubits radiated ($\log_2 R$). Correlations between the reference
(ref) subsystem and: (a) black hole interior, $B$; (b) radiation, $R$,
and external ($\text{ext}$) neighborhood; (c) black hole interior
and external neighborhood; and (d) radiation alone. Note that, as
expected from Eq.~(\ref{monogamy}), the sum of $C$'s in subplots (a)
and (b) is a constant, as is that of subplots (c) and (d). In each
subplot, the in-fallen matter consists of $S_{\text{matter}}= 10$ qubits and
the black hole initially consists of $\log_2 BR = 100$ qubits
with $\chi^{(q)}=0$. (Entropies are evaluated using base-two logarithms.)}
\label{results}
\end{figure}

For simplicity, here we restrict ourselves to the case where
\begin{equation}
\rho_{\text{ext}}=\frac{1}{M}\sum_{j=1}^M
 |j\rangle_{\text{ext}}\,{}_{\text{ext}}\!\langle j|,
\end{equation}
and where we assume no excess unentangled qubits, i.e., $\chi^{(q)}=0$.
Thus, the initial number of qubits within the black hole interior is
given by $\log_2 N =\log_2(BR)=S_{\text{matter}}+\log_2 M$, for
$S_{\text{matter}}$ qubits of in-fallen matter. We computed the above
measure of correlations, Eq.~(\ref{monogamy}), from von Neumann
entropies approximated using the average purity (see next section);
numerical calculations showed this as a good approximation
for systems of even a few qubits. Fig.~\ref{results} shows a typical
scenario: A black hole is assumed to be created from in-fallen matter
comprising $S_{\text{matter}}$ qubits of information and
negligible excess unentangled qubits. Within the first $S_{\text{matter}}$
qubits radiated, information about the in-fallen matter (a) vanishes
from the black hole interior at roughly the radiation emission rate
and (b) appears in the joint radiation and external neighborhood
subsystem. From then until just before the final $S_{\text{matter}}$
qubits are radiated, the in-fallen matter's information is encoded in
a tripartite state, involving the radiation, external neighborhood
and interior subsystems, subplots~(b) and~(c). In the final
$S_{\text{matter}}$ qubits radiated the information about the
in-fallen matter is released from its correlations and appears in the
radiation subsystem alone, subplot (d). This qualitative picture is
in excellent agreement with the results from the decoupling
theorem and its generalization.

\subsection*{Evaluation of purities}
\label{purities}

In order to approximate the computation of the correlation measure
described above, we use a lower bound for a subsystem with
density matrix $\rho$
\begin{equation}
\langle\!\langle S(\rho) \rangle\!\rangle
\ge -\langle\!\langle \,\log_2 p(\rho) \rangle\!\rangle
\ge -\log_2 \langle\!\langle p(\rho) \rangle\!\rangle. 
\end{equation}
Here $S(\rho)=-{\text{tr}}\, \rho \log_2 \rho$ is the von Neumann entropy
of $\rho$, $p(\rho)= {\text{tr}}\, \rho^2$ is its purity, and here
$\langle\!\langle \cdots \rangle\!\rangle$ denotes averaging over
random unitaries with the Haar measure. The former inequality above is
a consequence of the fact that the R\'enyi entropy is a non-increasing
function of its argument \cite{Bengtssonapp}, and the latter follows from the
concavity of the logarithm and Jensen's inequality. We may estimate
the von Neumann entropies required then by the rather crude
approximation
$\langle\!\langle S(\rho) \rangle\!\rangle
\approx -\log_2 \langle\!\langle \,p(\rho) \rangle\!\rangle$, 
which turns out to be quite reasonable for spaces with even a few qubits.

Although traditional methods \cite{Melloapp} may be used to compute these
purities, a much simpler approach is to use the approach from
Ref.~[\onlinecite{Abey06app}]. In particular, for a typical purity
of interest we use the following decomposition
\begin{eqnarray}
\text{tr}\; \sigma_{R,\text{ext}}^{U\;2}
&=& \text{tr} \bigl( \sigma_{R,\text{ext}}^U
\otimes \sigma_{R',\text{ext}'}^U \;
{\cal S}_{R,\text{ext};R',\text{ext}'}\bigr)\\
&=& \text{tr} \bigl( \rho_{\text{ref},BR,\text{ext}}\otimes
\rho_{\text{ref}',B'R',\text{ext}'}\nonumber\\
&&\times U_{BR}^\dagger\otimes U_{B'R'}^\dagger\,
{\cal S}_{R;R'}\, U_{BR}\otimes U_{B'R'}\,
{\cal S}_{\text{ext};\text{ext}'} \bigr) \nonumber
\end{eqnarray}
where ${\cal S}_{A;A'}$ is the swap operator between 
subsystems $A$ and $A'$, similarly,
${\cal S}_{AB;A'B'}={\cal S}_{A;A'}{\cal S}_{B;B'}$. Then the average
over the Haar measure is accomplished by an application of Schur's
lemma \cite{Abey06app}
\begin{eqnarray}
&&\bigl\langle\!\bigl\langle U_A^\dagger\otimes U_{A'}^\dagger\;
{\cal S}_{A_2;A_2'}\; U_A \otimes U_{A'}\bigr\rangle\!\bigr\rangle
\nonumber \\
&=&\frac{A_2(A_1^2-1)}{A^2-1}\;\openone_{A;A'}
+\frac{A_1(A_2^2-1)}{A^2-1}\;{\cal S}_{A;A'}.
\end{eqnarray}
This approach allows us to straight-forwardly compute the required
purities as
\begin{eqnarray}
p(\text{ref})\!&=&\!\frac{1}{K},\quad
p(\text{ext})=\frac{1}{N},\quad
p(\text{ref,ext})=\frac{1}{KN},~~~~~ \nonumber \\
p(R)\!&=&\!\frac{1}{(BR)^2-1}
\Bigl( R(B^2-1)+\frac{B(R^2-1)}{KN}\Bigr), \label{HJS} \\ 
p(R,\text{ext})\!&=&\!\frac{1}{(BR)^2-1}
\Bigl( \frac{R(B^2-1)}{N}+\frac{B(R^2-1)}{K}\Bigr), \nonumber
\end{eqnarray}
with $p(B,\text{ext})$ and $p(B,\text{ext})$ given by the
above expressions under the exchange $R\leftrightarrow B$, similarly
the exchange $K\leftrightarrow N$ gives us expressions for
$p(\text{ref},R)$, etc.

\section{XI.\ Black holes versus lumps of coal}
\label{coal}

Bekenstein \cite{Bekapp} tells of a thought experiment he attributes to
Sidney Coleman: A cold piece of coal (initially in its ground state) is
illuminated by a laser beam. The system is thus prepared in a pure state
and radiates thermally after the laser is switched off. Eventually
the lump of coal returns to its initial state, so presumably the
radiation subsystem has merely encoded any information in the subtle
correlations between the individual thermal photons. Can this differ
from the overall behavior of a unitarily evaporating black hole?

Such a hot coal model of a black hole will correspond very closely to
the pure-state model of a black hole. As such, information about the state
of the laser beam that has heated up the coal will become accessible
from the radiation field shortly after half of the total number of thermal
photons (each carrying roughly one bit's worth of information) have
radiated away. This behavior, however, will be very different from the
entangled black hole analyzed in our manuscript. For such highly entangled
black holes there is another component of the system to include in the
dynamics: The entanglement across the boundary corresponding to the
event horizon. This entanglement is not merely static  as it would be across
a fixed boundary, but must itself escape from the black hole in order
for the boundary itself to shrink. As was uncovered in our manuscript, 
entangled black holes encode the information about the in-fallen matter
into a quantum one-time pad. The information is in principle accessible
from any two of three subsystems (the interior of the black hole, the
modes just external to the black hole but entangled with it across
the event horizon and the Hawking radiation itself) within a very short
time after the black hole begins to radiate. Once that encoding into
the quantum one-time pad has occurred, this information becomes
inaccessible from any one of these subsystems alone (and in particular
from the Hawking radiation).

Only when the quantum one-time pad becomes decoded will the full information
become accessible within the Hawking radiation. For a highly entangled
black hole, as shown in our manuscript, this occurs within the final
and vanishingly small fraction of the black hole's lifetime. Before
this time, the Hawking radiation is completely uncorrelated from the
information about the in-fallen matter. This behavior is therefore
very different from that of information return from a hot coal.

\vskip 0.1truein
\noindent
The authors gratefully acknowledge H.-J.\ Sommers's original calculation of
Eq.~(\ref{HJS}) and several fruitful discussions with him, Netta Cohen
and Manas Patra.


\begin{thebibliography}{16}


\bibitem{Hawking75} S.\ W.\ Hawking,
Commun.\ Math.\ Phys.\ {\bf 43}, 199 (1975).

\bibitem{Nikolic} H.\ Nikoli\'c,
Int.\ J.\ Mod.\ Phys.\ D {\bf 14}, 2257 (2005);
S.\ D.\ Mathur,
Classical Quantum Gravity {\bf 26}, 224001 (2009).

\bibitem{B10} S.\ L.\ Braunstein and M.\ K.\ Patra,
Phys.\ Rev.\ Lett.\ {\bf 107}, 071302 (2011).



\bibitem{Hawking76} S.\ W.\ Hawking,
Phys.\ Rev.\ D {\bf 14}, 2460 (1976).

\bibitem{Page93} D.\ N.\ Page,
Phys.\ Rev.\ Lett.\ {\bf 71}, 3743 (1993).

\bibitem{Hayden07} P.\ Hayden and J.\ Preskill,
J.\ High Energy Phys.\ 09 (2007) 120.

\bibitem{Eisert09} J.\ Eisert, M.\ Cramer, and M.\ B.\ Plenio,
Rev.\ Mod.\ Phys.\ {\bf 82}, 277 (2010).

\bibitem{Smolin} J.\ A.\ Smolin and J.\ Oppenheim,
Phys.\ Rev.\ Lett.\ {\bf 96}, 081302 (2006).

\bibitem{me} S.\ L.\ Braunstein and A.\ K.\ Pati,
Phys.\ Rev.\ Lett.\ {\bf 98}, 080502 (2007).

\bibitem{Susskind93} L.\ Susskind, L.\ Thorlacius, and J.\ Uglum,
Phys.\ Rev.\ D {\bf 48}, 3743 (1993).

\bibitem{tHooft93} G.\ 't Hooft,
in {\it Salamfestschrift: A Collection of Talks}, 
edited by A.\ Ali, J.\ Ellis, and S.\ Randjbar-Daemi
(World Scientific, Singapore, 1993), Vol.\ 4.

\bibitem{Levy} V.\ Milman and G.\ Schechtman,
{\it Asymptotic Theory of Finite Dimensional Normed Spaces\/}
(Springer, New York, 2001).

\bibitem{Mehta} See, e.g., M.\ L.\ Mehta,
{\it Random Matrices\/}
(Elsevier, Amsterdam, 2004).

\bibitem{Abey06} A.\ Abeyesinghe, 
I.\ Devetak, P.\ Hayden, and A.\ Winter,
Proc.\ R.\ Soc.\ A {\bf 465}, 2537 (2009). 

\bibitem{SM} See Supplemental Material at
http://link.aps.org/\\
supplemental/10.1103/PhysRevLett.110.101301
for\\
proofs and a discussion of decoupling theorems.
(This material is included in full as the appendix.)

\bibitem{Schlieder} S.\ Schlieder,
Commun.\ Math.\ Phys.\ {\bf 1}, 256 (1965).

\bibitem{SLB2009} S.\ L.\ Braunstein, arXiv:0907.1190v1.

\bibitem{fnA} Our analysis is silent as to whether or not this
relationship between thermodynamic and entanglement entropy also
holds for acceleration or cosmological horizons.


\bibitem{bleach} J.\ Preskill,
in {\it Black Holes, Membranes, Wormholes and Superstrings},
edited by S.\ Kalara and D.\ V.\ Nanopoulos
(Singapore, World Scientific, 1993) p.~22.

\bibitem{fnB} All incoming degrees of freedom are required to be scrambled
for our information retrieval mechanism to apply in full. For black
holes with nontrivial interiors,
WKB tunneling analysis attributes a temperature to inner horizons
that always exceeds that of the outer
(event) horizon. Thus, the existence of an inner horizon is no absolute
barrier to global scrambling.

\bibitem{Sekino08} Y.\ Sekino and L.\ Susskind,
J.\ High Energy Phys.\ 10 (2008) 065.

\bibitem{Giddings07} S.\ B.\ Giddings,
Phys.\ Rev.\ D {\bf 76}, 064027 (2007).

\bibitem{Kretschmann} D.\ Kretschmann, D.\ Schlingemann, and R.\ F.\ Werner,
J.\ Funct.\ Anal.\ {\bf 255}, 1889 (2008);
J.\ R.\ Samal, A.\ K.\ Pati, and A.\ Kumar,
Phys.\ Rev.\ Lett.\ {\bf 106}, 080401 (2011).

\bibitem{tHooft85} G.\ 't Hooft,
Nucl.\ Phys.\ {\bf B256}, 727 (1985);
%
L.\ Bombelli, 
R.\ K.\ Koul, J.\ Lee, and R.\ D.\ Sorkin,
Phys.\ Rev.\ D {\bf 34}, 373 (1986);
%
M.\ Srednicki,
Phys.\ Rev.\ Lett.\ {\bf 71}, 666 (1993).

\bibitem{Hawking01}
S.\ Hawking, J.\ Maldacena, and A.\ Strominger,
J.\ High Energy Phys.\ 05 (2001) 001.

\bibitem{Brustein06}
R.\ Brustein, M.\ B.\ Einhorn, and A.\ Yarom,
J.\ High Energy Phys.\ 01 (2006) 098.
%
\bibitem{Emparan06}
R.\ Emparan,
J.\ High Energy Phys.\ 06 (2006) 012.

\bibitem{Pad10} T.\ Padmanabhan,
Phys.\ Rev.\ D {\bf 82}, 124025 (2010).


\bibitem{Lowe95} D.\ A.\ Lowe,
J.\ Polchinski, L.\ Susskind, L.\ Thorlacius, and J.\ Uglum,
Phys.\ Rev.\ D {\bf 52}, 6997 (1995).


\bibitem{Nishioka09} T.\ Nishioka, S.\ Ryu, and T.\ Takayanagi,
J.\ Phys.\ A {\bf 42}, 504008 (2009).

\bibitem{Brout} R.\ Brout,
Int.\ J.\ Mod.\ Phys.\ D {\bf 17}, 2549 (2008).

\bibitem{firewall} A.\ Almheiri, 
D.\ Marolf, J.\ Polchinski, and J.\ Sully,
J.\ High Energy Phys.\ 02 (2013) 062.

\vskip 0.2in
$~~~~~~~${\bf Supplemental Material bibliography}
\vskip 0.1in

\bibitem[$1_{\text{SM}}$]{Hayden07app}
See Ref.\ [\onlinecite{Hayden07}].

\bibitem[$2_{\text{SM}}$]{Susskind93app}
See Ref.\ [\onlinecite{Susskind93}].

\bibitem[$3_{\text{SM}}$]{Srednickiapp}
See Srednicki, Ref.\ [\onlinecite{tHooft85}].

\bibitem[$4_{\text{SM}}$]{Abey06app}
See Ref.\ [\onlinecite{Abey06}].

\bibitem[$5_{\text{SM}}$]{Page93app}
See Ref.\ [\onlinecite{Page93}].

\bibitem[$6_{\text{SM}}$]{B10app}
See Ref.\ [\onlinecite{B10}].

\bibitem[$7_{\text{SM}}$]{Bennett92app}
C.\ H.\ Bennett and S.\ J.\ Wiesner,
Phys.\ Rev.\ Lett.\ {\bf 69}, 2881 (1992).

\bibitem[$8_{\text{SM}}$]{meapp}
See Ref.\ [\onlinecite{me}].

\bibitem[$9_{\text{SM}}$]{Kretschmannapp}
See Ref.\ [\onlinecite{Kretschmann}].

\bibitem[$10_{\text{SM}}$]{Leung02app}
D.\ W.\ Leung,
Quantum Inf.\ Comput.\ {\bf 2}, 14 (2001).

\bibitem[$11_{\text{SM}}$]{Bengtssonapp}
I.\ Bengtsson and K.\ \.{Z}yczkowski,
Geometry of Quantum States: An Introduction to Quantum Entanglement.
(Cambridge University Press, Cambridge, 2006).

\bibitem[$12_{\text{SM}}$]{Melloapp} P.\ A.\ Mello,
J. Phys. A {\bf 23}, 4061 (1990).

\bibitem[$13_{\text{SM}}$]{Bekapp} J.\ D.\ Bekenstein,
Contemp.\ Phys.\ {\bf 45}, 31 (2004).

\end{thebibliography}
\end{document}